# Bounds and Computation of Irregularity of a Graph

November 8, 2018

**Hosam Abdo[a], Nathann Cohen[b], Darko Dimitrov[a]**

[a]*Institut für Informatik, Freie Universität Berlin,*
*Takustraße 9, D–14195 Berlin, Germany*
E-mail: `[abdo,darko]@mi.fu-berlin.de`

[b]*Algorithms Research Group, Computer Science Dept.*
*Université Libre de Bruxelles, 1050 Brussels, Belgium*
E-mail: `nathann.cohen@gmail.com`

**Abstract**

Albertson [3] has defined the *irregularity* of a simple undirected graph $G = (V, E)$ as $\mathrm{irr}(G) = \sum_{uv \in E} |d_G(u) - d_G(v)|$, where $d_G(u)$ denotes the degree of a vertex $u \in V$. Recently, this graph invariant gained interest in the chemical graph theory, where it occured in some bounds on the first and the second Zagreb index, and was named the *third Zagreb index* [13]. For general graphs with $n$ vertices, Albertson has obtained an asymptotically tight upper bound on the irregularity of $4n^3/27$. Here, by exploiting a different approach than in [3], we show that for general graphs with $n$ vertices the upper bound $\lfloor \frac{n}{3} \rfloor \lceil \frac{2n}{3} \rceil \left( \lceil \frac{2n}{3} \rceil - 1 \right)$ is sharp. Next, we determine $k$-cyclic graphs with maximal irregularity. We also present some bounds on the maximal/minimal irregularity of graphs with fixed minimal and/or maximal vertex degrees, and consider an approximate computation of the irregularity of a graph.

## 1 Introduction

Let $G = (V, E)$ be a simple undirected graph of order $n = |V|$ and size $m = |E|$. For $v \in V(G)$, the degree of $v$, denoted by $d_G(v)$, is the number of edges incident to $v$. Albertson [3] defines the *imbalance* of an edge $e = uv \in E$ as $|d_G(u) - d_G(v)|$ and the *irregularity* of $G$ as

$$\mathrm{irr}(G) = \sum_{uv \in E} |d_G(u) - d_G(v)|. \tag{1}$$

The *first Zagreb index* $M_1(G)$ and the *second Zagreb index* $M_2(G)$, are one of the oldest and most investigated topological graph indices, and are defined as follows:

$$\begin{aligned} M_1(G) &= \sum_{v \in V} d_G(v)^2, \\ M_2(G) &= \sum_{uv \in E} d_G(u) d_G(v). \end{aligned}$$

For details of the mathematical theory and chemical applications of the Zagreb indices see surveys [10, 15, 22, 28] and papers [11, 13, 30, 31, 32].

Recently in [13], Fath-Tabar established new bounds on the first and the second Zagreb indices which depend on the sum in (1). In line with the standard terminology of chemical graph theory, and the obvious connection with the first and the second Zagreb indices, Fath-Tabar named the sum in (1) the *third Zagreb index* and denoted it by $M_3(G)$. However, in the rest of the paper, we will use its older name and call it the irregularity of a graph.

Obviously, a connected graph $G$ has irregularity zero if and only if $G$ is regular. Other approaches, that characterize how irregular a graph is, have been proposed [1, 2, 6, 7, 8, 9, 16]. In this paper, we focus on graphs with maximal irregularity as defined in (1).

In [3] Albertson presented upper bounds on irregularity for bipartite graphs, triangle-free graphs and arbitrary graphs, as well as a sharp upper bound for trees. Some claims about bipartite graphs given in [3] have been formally proved in [20]. Related to Albertson [3] is the work of Hansen and Mélot [18], who characterized the graphs with $n$ vertices and $m$ edges with maximal irregularity. For more results on irregularity, imbalance, and related measures, we redirect the reader to [2, 4, 5, 23, 24, 25].

In the sequel we introduce the notation used in the rest of the paper.

The *maximal* and *minimal* degrees of a graph $G$ are denoted by $\Delta = \Delta(G)$ and $\delta = \delta(G)$, respectively. A *regular* graph is a graph where all the vertices have the same degree. A *pendant* vertex is a vertex of degree one. A *universal* vertex is the vertex adjacent to all other vertices. The *diameter* of a graph $G$ is the maximal distance between any two vertices of $G$. A set of vertices is said to be *independent* when the vertices are pairwise non-adjacent. The vertices from an independent set are *independent vertices*. By $N_G(u)$, we denote the set of vertices that are adjacent to a vertex $u$.

A *clique* of a graph $G$ is a complete subgraph of $G$. The *union* $G = G_1 \cup G_2$ of graphs $G_1$ and $G_2$ with disjoint vertex sets $V_1$ and $V_2$ and edge sets $E_1$ and $E_2$ is the graph with the vertex set $V = V_1 \cup V_2$ and the edge set $E = E_1 \cup E_2$. The *join* $G = G_1 + G_2$ of the graphs $G_1$ and $G_2$ is the graph union $G = G_1 \cup G_2$ together with all the edges joining $V_1$ and $V_2$.

The *clique-star* graph $KS_{p,q}$ is the join graph of a clique of size $p$ and an independent set of size $q$ (see Fig.1).

A sequence of non-negative integers $d_1, ..., d_n$ is a *graphic sequence*, or a *degree sequence*, if there exists a graph $G$ with $V(G) = \{v_1, ..., v_n\}$ such that $d(v_i) = d_i$. For characterizations and details of graphic sequences, we redirect an interested reader to [12, 14, 17, 19, 29].

## 2 General graphs with maximal irregularity

In order to characterize graphs with maximal irregularity, we first determine the minimum number of universal vertices that such graphs must have.

**Lemma 2.1.** *Let $G$ be a graph with maximal irregularity among all graphs of order n. Then, $G$ has at least $\lfloor \frac{n}{3} \rfloor$ universal vertices.*

*Proof.* Assume that $G$ is a graph with maximal irregularity whose set $U$ of universal vertices has cardinality $q < \lfloor \frac{n}{3} \rfloor$. Let $\overline{U} = \{\overline{u}_1, ..., \overline{u}_{n-q}\}$ be the set of non-universal vertices, where $d(\overline{u}_1) \geq d(\overline{u}_2) \geq \cdots \geq d(\overline{u}_{n-q-1}) \geq d(\overline{u}_{n-q})$.

If a non-neighbor $x$ of $\overline{u}_1$ is adjacent to a vertex $y \in \overline{U} \cap N_G(\overline{u}_1)$, then replace the edge $xy$ with the edge $\overline{u}_1 x$, obtaining a graph $G'$. By this replacement, the number of edges remains the same, as well as the degree of $x$. Also, the contribution of the edges between the $q$ universal vertices and the vertices of $\overline{U}$ to irr($G'$) remains unchanged.

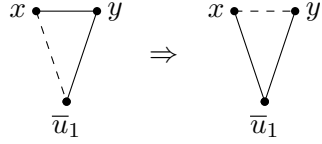

The contribution of the edge $xy$ to irr($G$) is $|d_G(y) - d_G(x)|$, and the contribution of the edge $\overline{u}_1 x$ to irr($G'$) is $d_G(\overline{u}_1) + 1 - d_G(x)$. After the above edge replacement, edges between $\overline{u}_1$ and $\overline{U} \cap N_G(\overline{u}_1)$ increase the irregularity by $d_G(\overline{u}_1) - q$, and edges between $y$ and $\overline{U} \setminus \{\overline{u}_1, x\}$ decrease the irregularity by at most $d_G(y) - q - 2$. Thus,

$$\begin{aligned}
\text{irr}(G') &= \text{irr}(G) - \overbrace{|d_G(y) - d_G(x)|}^{xy} + \overbrace{d_G(\overline{u}_1) + 1 - d_G(x)}^{\overline{u}_1 x} \\
&\quad + \underbrace{d_G(\overline{u}_1) - q}_{\text{edges from } \overline{u}_1 \text{ to } \overline{U} \cap N_G(\overline{u}_1)} \underbrace{-d_G(y) + q + 2}_{\text{edges from } y \text{ to } \overline{U} \setminus \{\overline{u}_1, x\}} \\
&= \text{irr}(G) \underbrace{-|d_G(y) - d_G(x)| - d_G(x) - d_G(y)}_{-2 \max(d_G(x), d_G(y))} + 2 d_G(\overline{u}_1) + 3 \\
&= \text{irr}(G) + 2\Big(d_G(\overline{u}_1) - \max(d_G(x), d_G(y))\Big) + 3.
\end{aligned}$$

Since $d_G(\overline{u}_1) \geq d_G(x), d_G(y)$, it follows that

$$\text{irr}(G') > \text{irr}(G).$$

We apply the above kind of replacement for all edges between $\overline{U} \cap N_G(\overline{u}_1)$ and $\overline{U} \setminus N_G(\overline{u}_1)$ – which only increases the irregularity of $G$ – so that we can now assume that there are none. Note also that $\overline{u}_1$ cannot have become an universal vertex, as it would contradict the assumption that a graph with maximal irregularity has at most $q$ universal vertices. Therefore, $\overline{u}_1$ is still the vertex of $\overline{U}$ of maximal degree. We denote by $G_1$ the newly obtained graph.

Next, we replace any edge $xy$ between two vertices $x, y \in \overline{U} \setminus N_{G_1}(\overline{u}_1)$ by $\overline{u}_1 x$ – this replacement preserves the number of edges as well the degree of $x$. The newly obtained graph we denote by $G''$. The contribution of the edges between the $q$ universal vertices and the vertices of $\overline{U}$ to irr($G_1$) are unchanged.

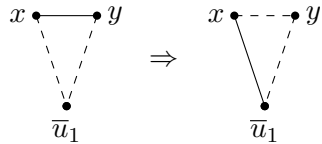

The contribution of edge $xy$ to irr($G_1$) is $|d_{G_1}(y) - d_{G_1}(x)|$, and the contribution of the edge $\overline{u}_1 x$ to irr($G''$) is $d_{G_1}(\overline{u}_1) + 1 - d_{G_1}(x)$. Edges between $\overline{u}_1$ and $\overline{U} \cap N_G(\overline{u}_1)$ increase the irregularity by $d_{G_1}(\overline{u}_1) - q$, and edges between $y$ and $\overline{U} \setminus N_G(\overline{u}_1) \setminus \{u\}$ decrease the irregularity by at most $d_{G_1}(y) - q - 1$.

Therefore,

$$\begin{aligned}
\mathrm{irr}(G'') &\geq \mathrm{irr}(G_1) - \overbrace{|d_{G_1}(y) - d_{G_1}(x)|}^{xy} + \overbrace{d_{G_1}(\overline{u}_1) + 1 - d_{G_1}(x)}^{\overline{u}_1 x} \\
&\quad + \underbrace{d_{G_1}(\overline{u}_1) - q}_{\text{edges from } \overline{u}_1 \text{ to } \overline{U} \cap N_{G_1}(\overline{u}_1)} \underbrace{-d_{G_1}(y) + q + 1}_{\text{edges from } y \text{ to } \overline{U} \setminus N_{G_1}(\overline{u}_1) \setminus \{x\}} \\
&= \mathrm{irr}(G_1) \underbrace{-|d_{G_1}(y) - d_{G_1}(x)| - d_{G_1}(y) - d_{G_1}(x)}_{-2\max(d_{G_1}(y), d_{G_1}(x))} + 2d_{G_1}(\overline{u}_1) + 2 \\
&= \mathrm{irr}(G_1) + 2(d_{G_1}(\overline{u}_1) - \max(d_{G_1}(y), d_{G_1}(x)) + 2.
\end{aligned}$$

As $d_{G_1}(\overline{u}_1) \geq d_{G_1}(y), d_{G_1}(x)$, we have that

$$\mathrm{irr}(G'') > \mathrm{irr}(G_1).$$

Hence, we can apply this second replacement to all edges between vertices of $\overline{U} \setminus N_{G_1}(\overline{u}_1)$ – which only increases $\mathrm{irr}(G_1)$ – and so assume that there are none. We denote by $G_2$ the newly obtained graph. As previously, $\overline{u}_1$ cannot become universal because of this procedure – that would contradict our assumptions on $G$.

Thus, $\overline{U} \setminus N_G(\overline{u}_1)$ is a nonempty independent set whose cardinality we note $z > 0$.

We can build a new graph $G^\star$ with $q + 1$ universal vertices from $G_2$, by linking $\overline{u}_1$ to its $z$ non-neighbor. As this operation changes the degree of $z + 1$ vertices, the contribution of the edges between the $q$ universal vertices and the rest of the vertices to $\mathrm{irr}(G^\star)$ is by $2zq$ smaller than their contribution to $\mathrm{irr}(G_2)$. However, The $z$ new edges between $\overline{u}_1$ and $\overline{U} \cap N_{G_2}(\overline{u}_1)$ contributes $z(n - 1 - q - 1)$ to $\mathrm{irr}(G^\star)$, and the contribution between $\overline{u}_1$ and vertices of $\overline{U} \cap N_{G_2}(\overline{u}_1)$ increases in $\mathrm{irr}(G^\star)$ by $z(n - q - z - 1)$. Therefore,

$$\begin{aligned}
\mathrm{irr}(G^\star) &= \mathrm{irr}(G_2) - 2zq + z(n - 1 - q - 1) + z(n - q - z - 1) \\
&= \mathrm{irr}(G_2) + z(2n - 4q - z - 3).
\end{aligned} \qquad (2)$$

As $z \leq n - q - 1$, further we have

$$\mathrm{irr}(G^\star) \geq \mathrm{irr}(G_2) + z(n - 3q - 2). \qquad (3)$$

Since we have assumed $q < \lfloor \frac{n}{3} \rfloor$, it follows that $\mathrm{irr}(G^\star) > \mathrm{irr}(G_2)$. Thus, we have shown that we can obtain a graph $G^\star$ with $q + 1$ universal vertices with irregularity greater than any graph $G$ with maximal irregularity, which is a contradiction to the assumption that any graph $G$ with maximal irregularity has at most $q$ universal vertices. □

We will now determine the graphs whose irregularity is maximum.

**Theorem 2.1.** *If a graph $G$ has maximal irregularity among all graphs of order $n$, then $G$ is either the clique-star graph $KS_{\lfloor \frac{n}{3} \rfloor, \lceil \frac{2n}{3} \rceil}$, or, if $n \equiv 2 \pmod{3}$, the clique-star graph $KS_{\lceil \frac{n}{3} \rceil, \lfloor \frac{2n}{3} \rfloor}$.*

*Proof.* Let $G$ be a graph of maximum irregularity, and let $U = \{v_{n-q+1}, v_{n-q+2}, \ldots, v_n\}$ be the set of universal vertices, where $q \geq \lfloor \frac{n}{3} \rfloor$ (cf. Lemma 2.1). Let $\overline{U}$ be the set of non-universal vertices, let $G[\overline{U}]$ be the graph induced by all non-universal vertices, and let $G' = G - G[\overline{U}]$ be the complement of $G[\overline{U}]$ in $G$.

As $d_{G'}(v) = d_G(v) - d_{\overline{U}}(v) = q$, the edges between $U$ and $\overline{U}$ contribute more to irr($G'$) than they do to irr($G$), by a difference of

$$\sum_{v \in \overline{U}} d_{\overline{U}}(v) q. \qquad (4)$$

On the other hand, the contribution of edges from $G[\overline{U}]$ does not appear in the computation of irr($G'$). The difference of the degrees between the endvertices of an edge of $G[\overline{U}]$ is at most $n-q-3$. Therefore, the edges from $G[\overline{U}]$ contribute to irr($G$) by most

$$\frac{1}{2} \sum_{v \in \overline{U}} d_{\overline{U}}(v) (n - q - 3). \qquad (5)$$

From (4) and (5), we have

$$\begin{aligned} \text{irr}(G') &\geq \text{irr}(G) + \sum_{v \in \overline{U}} d_{\overline{U}}(v) q - \frac{1}{2} \sum_{v \in \overline{U}} d_{\overline{U}}(v)(n - q - 3) \\ &= \text{irr}(G) + \left( \frac{1}{2}(3q - n) + \frac{3}{2} \right) \sum_{v \in \overline{U}} d_{\overline{U}}(v). \end{aligned}$$

The expression $\frac{1}{2}(3q - n) + \frac{3}{2}$ is positive for $q \geq \lfloor \frac{n}{3} \rfloor$. Since, $G$ is a graph with maximal irregularity, it follows that $\sum_{v \in \overline{U}} d_{\overline{U}}(v) = 0$, i.e., the vertices of $\overline{U}$ form an independent set. Therefore, $G$ is a clique-star graph $KS_{q,n-q}$, with $q \geq \lfloor \frac{n}{3} \rfloor$. The irregularity of $KS_{q,n-q}$ is $q(n-q)(n-1-q)$, and it is maximized for $q = \lfloor \frac{n}{3} \rfloor$, and for $q = \lceil \frac{n}{3} \rceil$, if $n \equiv 2 \pmod 3$. □

The graphs with maximal irregularity with 6, 7 and 8 vertices are depicted in Figure 1.

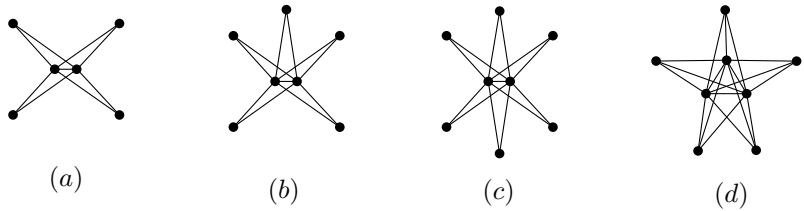

$(a) \qquad (b) \qquad (c) \qquad (d)$

Figure 1: (a) The graph with 6 vertices with maximal irr. (b) The graph with 7 vertices with maximal irr. (c) and (d) Graphs with 8 vertices with maximal irr.

**Corollary 2.1.** *For any $G$, $\text{irr}(G) \leq \lfloor \frac{n}{3} \rfloor \lceil \frac{2n}{3} \rceil \left( \lceil \frac{2n}{3} \rceil - 1 \right) = \text{irr}(KS_{\lfloor \frac{n}{3} \rfloor, \lceil \frac{2n}{3} \rceil}).$*

## 3 $k$-cyclic graphs with maximal irregularity

In this section we present graphs with maximal irregularity among $k-$cyclic graphs with $n$ vertices.

First, we make the following observation: for a connected $k-$cyclic graphs with $n$ vertices and $m$ edges, it holds that $k = m - n + 1$. Thus, connected graphs with same number of vertices and edges have same cyclomatic number. Also, the proof of Theorem 3.1 can be seen as alternative proof for the problem of determining a graph with maximal irr among all graphs with given $n$ and $m$, which was accomplished by Hansen and Mélot in [18].

Before, we present the main result in this section, we refer to the definition of a fanned split graph given in [18]. A *fanned split* graph $FS_{n_u n_1}$ is a graph comprised of $n_u$ universal vertices, a vertex $v$ of degree $n_u + n_1$, $n_1$ vertices adjacent to the universal vertices and the vertex $v$, and additional $n - n_u - n_1 - 1$ independent vertices adjacent only to the universal vertices. A fanned split graph can also be thought of as a clique-star graph to which have been added several edges by picking one non-universal vertex $v$ and making it adjacent to other non-universal vertices A straightforward calculation gives that

$$\text{irr}(FS_{n_u n_1}) = n_u(n - n_u)(n - n_u - 1) + n_1(n_1 - 2n_u - 1).$$

The number of edges adjacent to the universal vertices of $FS_{n_u n_1}$ is $n_u(n_u - 1)/2 + n_u(n - n_u)$. From the description of $FS_{n_u n_1}$, it follows that $FS_{n_u n_1}$ has the maximal number of universal vertices among all graphs with same number of vertices and edges, and therefore, $n_u$ is the largest integer that satisfies $n_u(n_u - 1)/2 + n_u(n - n_u) \leq m = n + k - 1$. From here, we have that

$$n_u = \left\lfloor \frac{1}{2}\left(2n - 1 - \sqrt{(2n-3)^2 - 8k}\right)\right\rfloor$$

and

$$n_1 = n + k + \frac{1}{2}n_u(2n - 3n_u + 1) - 1.$$

**Theorem 3.1.** *The graph with maximal irregularity among all graphs k-cyclic graphs of same order is a fanned split graph.*

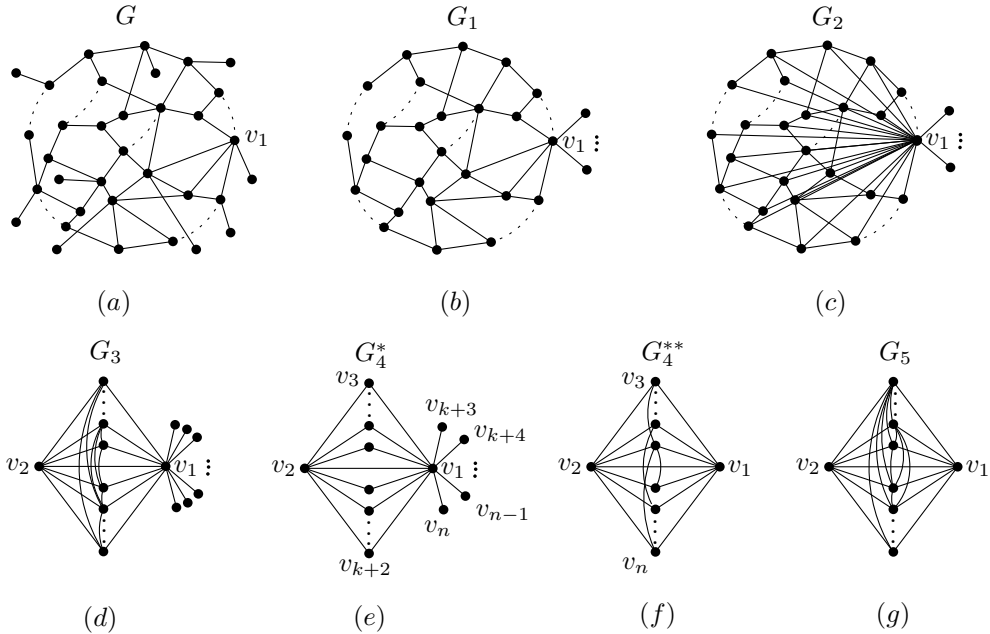

Figure 2: Seven $k-$cyclic graphs that illustrate the proof of Theorem 3.1.

*Proof.* Let $G$ be a $k-$cyclic graph $G$ of order $n$. We prove the theorem in five steps. In each step we show how $G$ can be modified so that its irregularity increases at constant number of edges, to finally deduce that a graph with maximal irregularity is a fanned split graph.

**Step** 1. *Attaching all pendant vertices, if any, to the vertex with maximal degree.*

Let $v_1$ be a vertex with maximal degree, and let us suppose that there exist pendant vertices nonadjacent to $v_1$. If such vertices do not exist, this step can be skipped.

Let $G_1$ be the graph in which all pendant vertices that are not adjacent to $v_1$ have been removed, and replaced by as many new pendant vertices adjacent to $v_1$. We have $\text{irr}(G_1) > \text{irr}(G)$ as the replacement of any pendant vertex $p$ adjacent to $v$ by a pendant vertex adjacent to $v_1$ produces an increase of irr$(G)$ by at least

$$d_G(v_1) + d_G(v_1) - (d_G(z) - 1) - (d_G(z) - 1) > 0. \tag{6}$$

It follows that irr$(G_1) >$ irr$(G)$.

**Step** 2. *Making the vertex with maximal degree an universal vertex.*

If the vertex $v_1$ is universal, skip this step.

Denote by $V_n(v_1)$ the set of all vertices in $G_1$ that are not adjacent to $v_1$. Observe that every vertex in $V_n(v_1)$ has at least degree two. Choose an edge $uv$ such that only the endvertex $u$ is in $V_n(v_1)$. Notice that such an edge must exist as otherwise $G$ is not connected. Delete the edge $uv$, and add the edge $uv_1$, obtaining a graph $G'_2$. After this replacement, the irregularity of $G_1$ increases at least by

$$(d_{G_1}(v_1) + 1 - d_{G_1}(u)) + d_{G_1}(v_1) - |d_{G_1}(v) - d_{G_1}(u)| - (d_{G_1}(v) - 2). \tag{7}$$

The irregularity of the graph $G_1$ increases at least by

$$2(d_{G_1}(v_1) - \max\{d_{G_1}(v), d_{G_1}(u)\}) + 3 > 0. \tag{8}$$

It follows that irr$(G'_2) >$ irr$(G_1)$. $G'_2$ is also $k-$cyclic. Repeat this kind of replacement for all edges with exactly one endvertex in $V_n(v_1)$ obtaining a graph $G_2$ with irr$(G_2) >$ irr$(G'_2)$.

**Step** 3. *Connect all possible vertices (except the pendant ones) to the vertex with second maximal degree.*

Denote by $v_2$ the vertex with the second maximal degree in $G_2$ and by $E_n(v_2)$ the set of edges that have at least one endvertex not adjacent to $v_2$. Let, for an edge $xy$ in $E_n(v_2)$, $x$ be a vertex that is not adjacent to $v_2$. Replace the edge $xy$ with the edge $xv_2$ obtaining a graph $G'_3$. Then, the irregularity of the graph $G_2$ increases at least by

$$(d_{G_2}(v_2) + 1 - d_{G_2}(x)) + d_{G_2}(v_2) - 1 - |d_{G_2}(x) - d_{G_2}(y)| - (d_{G_2}(y) - 2). \tag{9}$$

The irregularity of the graph $G_1$ increases at least by

$$2(d_{G_2}(v_2) - \max\{d_{G_2}(x), d_{G_2}(y)\}) + 2 > 0. \tag{10}$$

For as long as it is possible, do the above replacement to obtain a graph $G_3$ with irr$(G_3) >$ irr$(G'_3)$.

**Step** 4. *Making the pendant vertices, if any, adjacent to the vertex with second largest degree.*

If, after step 3, there is an edge $uv$ with both endvertices in $V_2 = V(G_3) \setminus \{v_1, v_2\}$, and there are pendant vertices in $V(G_3)$ (as illustrated in Figure 2(d)), then replace the edge $uv$ with the edge $xv_2$, where $x$ is a pendant vertex, obtaining a graph $G'_4$. Note that if there is no such edge, then $G$ is already a fanned split graph. It holds that

$$\begin{aligned}
\text{irr}(G'_4) &\geq \text{irr}(G_3) - |d_{G_3}(u) - d_{G_3}(v)| - (d_{G_3}(u) - 3) - (d_{G_3}(v) - 3) + \\
&\quad (d_{G_3}(v_2) - 1) + (d_{G_3}(v_2) + 1 - 2) - 1 \\
&= \text{irr}(G_3) - |d_{G_3}(u) - d_{G_3}(v)| - d_{G_3}(u) - d_{G_3}(v) + 2d_{G_3}(v_2) + 3 \\
&> \text{irr}(G_3). \tag{11}
\end{aligned}$$

Repeat this for all edges with both endvertices in $V_2$ that can be replaced by an edge between a pendant vertex and $v_2$, obtaining graph $G_4$ with irr$(G_4) >$ irr$(G_3)$. Observe that if $k \leq n - 2$, then $G_4$ has no edges between any two vertices of $V_2$, as the graph $G_4^*$ in Figure 2(e). In this case, any replacement of an edge (or sequence of edges) will result in a graph obtained during the first four steps, and therefore will have smaller irregularity than $G_4$. If $k \geq n - 2$, after this step, the graph $G_4$ does not have pendant vertices, as the graph $G_4^{**}$ in Figure 2(f).

**Step** 5. *Increase the degrees of the non-universal vertices, that have largest degrees.*

Because of steps 2, 3, and 4 we know that there are at least two vertices of maximum degree. Hence, let us assume that there are $m-1$ universal vertices, where $3 \leq m \leq n-3$. Let $V_{nu} = \{v \in G_4 |$ v is not an universal vertex$\}$. Let $v_m$ be the vertex with maximal degree among the set $V_{nu}$ of all non universal vertices. Let $u$ be a non-universal vertex different from $v_m$. Let $v$ be in $V_{nu} \setminus \{v_m\}$, that is adjacent to $u$. Replace $uv$ with $uv_3$, obtaining a graph $G'_5$. We have

$$\begin{aligned}
\mathrm{irr}(G'_5) &\geq \mathrm{irr}(G_4) - |d_{G_4}(u) - d_{G_4}(v)| - (d_{G_4}(v) - m - 2) + m + \\
&\quad (d_{G_4}(v_m) + 1 - d_{G_4}(u)) + (d_{G_4}(v_m) - m - 2) \\
&> \mathrm{irr}(G_4).
\end{aligned} \qquad (12)$$

Update $V_{nu}$ and $v_m$, and continue the above kind of replacement, i.e., obtaining a graph $G_5$ with $\mathrm{irr}(G_5) > \mathrm{irr}(G_4) > \mathrm{irr}(G_3) > \mathrm{irr}(G_2) > \mathrm{irr}(G_1) > \mathrm{irr}(G)$. Observe that $G_5$ is a fanned split graph. □

## 4 Bounds on graphs with maximal and minimal irregularity

### 4.1 Lower bounds on graphs with maximal irregularity

In this section, we consider graphs with maximal irregularity and prescribed minimal or/and maximal degrees. First, we show a lower bound for graphs with fixed maximal degree $\Delta$.

**Proposition 4.1.** *Let $G$ be a connected graph with n vertices with maximum degree $\Delta(G) = \Delta$, and maximal irregularity. Then, it holds that*

$$\mathrm{irr}(G) \geq \frac{4\Delta^2 n}{27} + nO(\Delta).$$

*Proof.* To obtain the bound we consider the graph $Q$ which is illustrated in Figure 3.

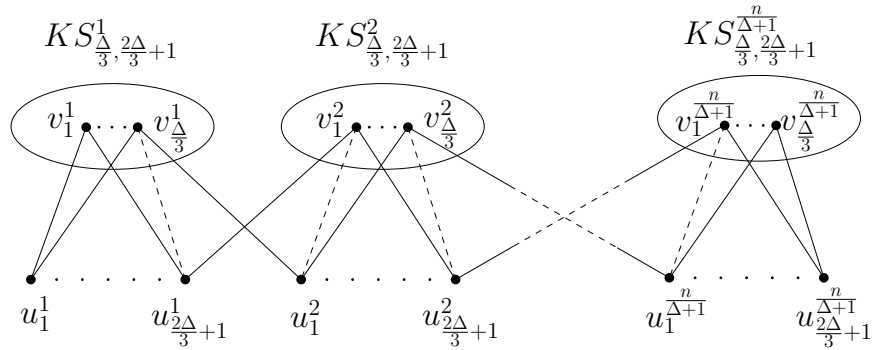

Figure 3: A connected graph $Q$ constructed from $n/(\Delta+1)$ copies of $KS_{\frac{\Delta}{3}, \frac{2\Delta}{3}+1}$. The dashed edges are those that are removed from the corresponding clique-star graphs.

To simplify the calculation, we assume that $\Delta/3$ and $n/(\Delta+1)$ are integers. The construction of $Q$ is as follows:

- Make a sequence of $n/(\Delta+1)$ copies of $KS_{\frac{\Delta}{3}, \frac{2\Delta}{3}+1}$.
- Choose an edge from the first $KS_{\frac{\Delta}{3}, \frac{2\Delta}{3}+1}$ graph, with an independent vertex as one endvertex, and an edge from the second $KS_{\frac{\Delta}{3}, \frac{2\Delta}{3}+1}$ graph, also with an independent vertex as one endvertex. Let denote these edges by $v^1_{\frac{\Delta}{3}} u^1_{\frac{2\Delta}{3}+1}$ and $v^2_1 u^2_1$, respectively. Replace $v^1_{\frac{\Delta}{3}} u^1_{\frac{2\Delta}{3}+1}$ and $v^2_1 u^2_1$ by edges $v^1_{\frac{\Delta}{3}} u^2_1$ and $v^2_1 u^1_{\frac{2\Delta}{3}+1}$. Continue this kind of replacement between all consecutive copies of $KS_{\frac{\Delta}{3}, \frac{2\Delta}{3}+1}$. Notice that these replacements do not change the degrees of the vertices.

We have
$$\operatorname{irr}(Q) = \frac{n}{\Delta+1}\frac{\Delta}{3}\left(\frac{2\Delta}{3}+1\right)\frac{2\Delta}{3} = \frac{4\Delta^2 n}{27}+nO(\Delta).$$
□

Next, we show a lower bound for graphs with maximal irregularity and fixed minimal degree $\delta$.

**Proposition 4.2.** *Let $G$ be a connected graph with $n$ vertices with minimal degree $\delta(G) = \delta$, and maximal irregularity. Then*
$$\operatorname{irr}(G) \geq \delta(n-\delta)(n-1-\delta) = \Omega(\delta n^2).$$

*Proof.* The lower bound is obtained by $KS_{\delta,n-\delta}$ whose irregularity is $\delta(n-\delta)(n-1-\delta)$. □

Finally, we show a lower bound for graphs with maximal irregularity and fixed maximal and minimal degrees.

**Proposition 4.3.** *Let $G$ be a connected graph with $n$ vertices with minimal degree $\delta(G) = \delta$, maximal degree $\Delta(G) = \Delta$, and maximal irregularity. Then, it holds*
$$\operatorname{irr}(G) > \frac{\delta}{\Delta+1}(\Delta-\delta)^2 n.$$

*Proof.* To obtain the bound we consider the graph $R$, which is constructed in the same way as the graph $Q$ in Figure 3, with only difference that $R$ is built of $n/(\Delta+1)$ copies of $KS_{\delta,\Delta-\delta+1}$. To simplify the calculation, we assume that $n/(\Delta+1)$ is integer. We have
$$\operatorname{irr}(R) = \frac{n}{\Delta+1}\delta(\Delta-\delta+1)(\Delta-\delta).$$
□

## 4.2 Upper bounds on graphs with minimal irregularity

In this section, we consider graphs with minimal irregularity and prescribed minimal and/or maximal degrees. To prove the results in this section, we will use the following characterization of graphic sequences.

**Theorem 4.1** (Erdős-Gallai [12]). *A sequence $d_1 \geq d_2 \geq \cdots \geq d_n$ of non-negative integers, whose sum is even is graphic if and only if*
$$\sum_{i=1}^{r} d_i \leq r(r-1) + \sum_{i=r+1}^{n} \min(r, d_i),$$
*for all $1 \leq r \leq n$.*

First, we show an upper bound for graphs with fixed maximal degree $\Delta$.

**Proposition 4.4.** *Let $G$ be a connected graph with $n$ vertices with maximal degree $\Delta(G) = \Delta$, and minimal irregularity. If $\Delta$ or $n$ is even, then $\operatorname{irr}(G) = 0$. If $\Delta$ and $n$ are odd, then $\operatorname{irr}(G) \leq \Delta - 1$.*

*Proof.* If $\Delta$ or $n$ is even, then by Theorem 4.1, there exist a $\Delta$-regular graph with $n$ vertices.
If $\Delta$ and $n$ are odd, then by Theorem 4.1, there exist a graph comprised of $n-1$ vertices of degree $\Delta$, and one vertex of degree $\Delta - 1$. The irregularity of this graph is $\Delta - 1$. □

Next, we show an upper bound for graphs with minimal irregularity and fixed minimal degree $\delta$.

**Proposition 4.5.** *Let $G$ be a connected graph with $n$ vertices with minimal degree $\delta(G) = \delta$, and minimal irregularity. If $\delta$ or $n$ is even, then $irr(G) = 0$. If $\delta$ and $n$ are odd, then $\mathrm{irr}(G) \leq \delta + 1$.*

*Proof.* If $\delta$ or $n$ is even, then by Theorem 4.1, there exist a $\delta$-regular graph with $n$ vertices.

If $\delta$ and $n$ are odd, then by Theorem 4.1, there exist a graph comprised of $n-1$ vertices of degree $\delta$, and one vertex of degree $\delta + 1$, whose irregularity is $\delta + 1$. □

Finally, we show an upper bound for graphs with minimal irregularity and fixed maximal and minimal degrees.

**Proposition 4.6.** *Let $G$ be a connected graph with $n$ vertices with minimal degree $\delta(G) = \delta$, maximal degree $\Delta(G) = \Delta$, and minimal irregularity. Then,*

$$\mathrm{irr}(G) \leq 2\Delta(\Delta - \delta).$$

*Proof.* First, assume that at least one of $n$, $\delta$ and $\Delta$ is even. If $\delta$ (resp. $\Delta$) is odd, consider the degree sequence comprised of two vertices of degree $\delta$ (resp. $\Delta$) and rest of the vertices of degree $\Delta$ (resp. $\delta$). By Theorem 4.1 there exists a graph with such degree sequence and has irregularity at most $2\delta(\Delta - \delta)$ (resp. $2\Delta(\Delta - \delta)$).

Second, let $n$, $\delta$ and $\Delta$ be odd. Then consider the degree sequence comprised of one vertex of degree $\delta + 1$, two vertices of degree $\Delta$ and the rest of the vertices of degree $\delta$. Theorem 4.1 guarantees the existence of a graph with such degree sequence and its irregularity is at most $2\Delta(\Delta - \delta)$. □

## 5 Exact and approximative computations of the irregularity of a graph

In order to better understand the properties of graphs with large irregularity, we thought sensible to approach the problem by enumerating the graphs with a fixed number $n$ of vertices in order to compute the irregularity of each of them. While such a procedure is made easy by the software Sage [27], or Brendan McKay's Nauty [21], the exhaustive enumeration of graphs quickly becomes impractical due to the sheer number of such graphs (which happens in practice as soon as $n \approx 11$). Therefore, we attempted to relax the computational problem by enumerating the possible degree sequences of graphs with a fixed number of vertices instead of the graphs themselves. Indeed, the number of different degree sequences of graphs with $n$ vertices is fairly small compared to the number of non-isomorphic graphs, and the code necessary to enumerate them much simpler[1].

One can not hope, however, to compute the value of irr with only a degree sequence, though it is possible to upper-bound the irregularity of a graph $G$ with this information. The following lines describe a bound on the irregularity of a graph depending only on its degree sequence.

$$\begin{aligned}\mathrm{irr}(G) &= \sum_{uv \in E(G)} |d(u) - d(v)| \\ &= \sum_{0 \leq i < n} |\{uv \in E(G) : d(u) \leq i \text{ and } d(v) > i\}|.\end{aligned}$$

Let us now write $d_{\leq i}$ (resp. $d_{>i}$) the number of vertices of $G$ whose degree is smaller (resp. strictly larger) than $i$. Given a vertex $v$ of degree $\leq i$, the number of neighbors of degree $> i$ it can have is necessarily smaller than both $d_{>i}$ and $d(v)$. Following the same steps for vertices $v$ of degree $> i$, we obtain

$$\mathrm{irr}(G) \leq \sum_{0 \leq i < n} \min\Big[\sum_{\substack{v \in V(G) \\ d(v) \leq i}} \min\big(d(v), d_{>i}\big), \sum_{\substack{v \in V(G) \\ d(v) > i}} \min\big(d(v), d_{\leq i}\big)\Big]. \qquad (13)$$

| $n$ | $KS_n^{max}$ | graphic sequences of $KS_n^{max}$ | $irr(KS_n^{max})$ |
|---|---|---|---|
| 3 | $KS_{1,2}$ | [2, 1, 1] | 2 |
| 4 | $KS_{1,3}$ | [3, 1, 1, 1] | 6 |
| 5 | $KS_{1,4}$ | [4, 1, 1, 1, 1] | 12 |
|   | $KS_{2,3}$ | [4, 4, 2, 2, 2] | 12 |
| 6 | $KS_{2,4}$ | [5, 5, 2, 2, 2, 2] | 24 |
| 7 | $KS_{2,5}$ | [6, 6, 2, 2, 2, 2, 2] | 40 |
| 8 | $KS_{2,6}$ | [7, 7, 2, 2, 2, 2, 2, 2] | 60 |
|   | $KS_{3,5}$ | [7, 7, 7, 3, 3, 3, 3, 3] | 60 |
| 9 | $KS_{3,6}$ | [8, 8, 8, 3, 3, 3, 3, 3, 3] | 90 |
| 10 | $KS_{3,7}$ | [9, 9, 9, 3, 3, 3, 3, 3, 3, 3] | 126 |
| 11 | $KS_{3,8}$ | [10, 10, 10, 3, 3, 3, 3, 3, 3, 3, 3] | 168 |
|   | $KS_{4,7}$ | [10, 10, 10, 10, 4, 4, 4, 4, 4, 4, 4] | 168 |
| 12 | $KS_{4,8}$ | [11, 11, 11, 11, 4, 4, 4, 4, 4, 4, 4, 4] | 224 |
| 13 | $KS_{4,9}$ | [12, 12, 12, 12, 4, 4, 4, 4, 4, 4, 4, 4, 4] | 288 |
| 14 | $KS_{4,10}$ | [13, 13, 13, 13, 4, 4, 4, 4, 4, 4, 4, 4, 4, 4] | 360 |
|   | $KS_{5,9}$ | [13, 13, 13, 13, 13, 5, 5, 5, 5, 5, 5, 5, 5, 5] | 360 |
| 15 | $KS_{5,10}$ | [14, 14, 14, 14, 14, 5, 5, 5, 5, 5, 5, 5, 5, 5, 5] | 450 |
| 16 | $KS_{5,11}$ | [15, 15, 15, 15, 15, 5, 5, 5, 5, 5, 5, 5, 5, 5, 5, 5] | 550 |
| 17 | $KS_{5,12}$ | [16, 16, 16, 16, 16, 5, 5, 5, 5, 5, 5, 5, 5, 5, 5, 5, 5] | 660 |
|   | $KS_{6,11}$ | [16, 16, 16, 16, 16, 16, 6, 6, 6, 6, 6, 6, 6, 6, 6, 6, 6] | 660 |
| 18 | $KS_{6,12}$ | [17, 17, 17, 17, 17, 17, 6, 6, 6, 6, 6, 6, 6, 6, 6, 6, 6, 6] | 792 |
| 19 | $KS_{6,13}$ | [18, 18, 18, 18, 18, 18, 6, 6, 6, 6, 6, 6, 6, 6, 6, 6, 6, 6, 6] | 936 |
| 20 | $KS_{6,14}$ | [19, 19, 19, 19, 19, 19, 6, 6, 6, 6, 6, 6, 6, 6, 6, 6, 6, 6, 6, 6] | 1092 |
|   | $KS_{7,13}$ | [19, 19, 19, 19, 19, 19, 19, 7, 7, 7, 7, 7, 7, 7, 7, 7, 7, 7, 7, 7] | 1092 |
| 21 | $KS_{7,14}$ | [20, 20, 20, 20, 20, 20, 20, 7, 7, 7, 7, 7, 7, 7, 7, 7, 7, 7, 7, 7, 7] | 1274 |
| 22 | $KS_{7,15}$ | [21, 21, 21, 21, 21, 21, 21, 7, 7, 7, 7, 7, 7, 7, 7, 7, 7, 7, 7, 7, 7, 7] | 1470 |
| 23 | $KS_{7,16}$ | [22, 22, 22, 22, 22, 22, 22, 7, 7, 7, 7, 7, 7, 7, 7, 7, 7, 7, 7, 7, 7, 7, 7] | 1680 |
|   | $KS_{8,15}$ | [22, 22, 22, 22, 22, 22, 22, 22, 8, 8, 8, 8, 8, 8, 8, 8, 8, 8, 8, 8, 8, 8, 8] | 1680 |

Table 1: Graphs with the maximal irregularity, their corresponding graphic sequences, and the values of their irregularities.

By computing this bound on all the degree sequences of graphs with $n$ vertices, we obtained the list of the degree sequences for which the value reached by this bound is maximal. This would not have necessarily meant that a graph having such degree sequence would have the largest irregularity among all graphs with $n$ vertices – for it is only an upper bound on the irregularity of such a graph – though we remarked in this situation that the degree sequences for which this bound was the largest corresponded to the graphs described by Theorem 2.1. In particular, as for these graphs the upper bound is equal to the irregularity, those graphs are indeed the (only) extremal ones.

In Table 1 are gathered the results of our experiments up to $n = 23$, where the number of degree sequences, in turn, became too large to continue further. It contains for each $n$ the degree sequences maximizing the bound (13), along with a corresponding graph for which irr is equal to the upper bound. $KS_n^{max}$ denotes a graph with the maximal irregularity among all clique-star graphs with $n$ vertices. By Theorem 2.1, $KS_n^{max}$ has the maximal irregularity among all graphs with $n$ vertices, and is determined by

$$KS_n^{max} = \begin{cases} KS_{\lfloor \frac{n}{3} \rfloor, \lceil \frac{2n}{3} \rceil} \text{ and } KS_{\lceil \frac{n}{3} \rceil, \lfloor \frac{2n}{3} \rfloor}, & \text{if } n \equiv 2 \pmod{3}, \\ KS_{\lfloor \frac{n}{3} \rfloor, \lceil \frac{2n}{3} \rceil}, & \text{otherwise.} \end{cases} \quad (14)$$

---
[1]Our implementation used the software Sage [27], and has been submitted for inclusion in the software. Its source code is available on its trac ticket [26].

# 6  Acknowledgment

We would like to thank Ivan Gutman for rising the question about graphs with maximal third Zagreb index and for providing us with useful references regarding the subject considered in this paper.